\documentclass[runningheads]{llncs}
\usepackage[T1]{fontenc}
\usepackage{graphicx}
\usepackage{float}
\usepackage{subcaption}
\usepackage{xspace}
\usepackage{arydshln}
\usepackage{multirow}
\usepackage{amsmath}
\usepackage[table]{xcolor}
\usepackage[framemethod=TikZ]{mdframed}
\newcolumntype{P}[1]{>{\centering\arraybackslash}p{#1}}
\newcommand{\GOD}{GOD\xspace}
\newcommand{\TOD}{TOD\xspace}
\newcommand{\ID}{ID\xspace}
\newcommand{\tool}{AI-DO\xspace} 
\newcommand{\ie}{\emph{i.e.,}\xspace}
\newcommand{\eg}{\emph{e.g.,}\xspace}

\newcommand{\etal}{\emph{et~al.}\xspace}
\newcommand{\secref}[1]{Section~\ref{#1}\xspace}

\newcommand{\figref}[1]{Fig.~\ref{#1}\xspace}

\newcommand{\tabref}[1]{Table~\ref{#1}\xspace}

\begin{document}

\title{Cross-Domain Evaluation of Transformer-Based Vulnerability Detection on Open \& Industry Data}

\author{
    Moritz Mock\inst{1}
    \and
    Thomas Forrer\inst{2} 
    \and
    Barbara Russo\inst{1}
}
\authorrunning{Mock \etal}
\titlerunning{Cross-Domain Evaluation}
\institute{Faculty of Engineering, Free University of Bozen-Bolzano, Bolzano, Italy \\ \email{\{momock,brusso\}@unibz.it} \and
R\&D Department, W{\"u}rth Phoenix, Bolzano, Italy
\email{thomas.forrer@wuerth-phoenix.net}}

\maketitle              

\begin{abstract}
Deep learning solutions for vulnerability detection proposed in academic research are not always accessible to developers, and their applicability in industrial settings is rarely addressed. 
Transferring such technologies from academia to industry presents challenges related to trustworthiness, legacy systems, limited digital literacy, and the gap between academic and industrial expertise. For deep learning in particular, performance and integration into existing workflows are additional concerns.
In this work, we first evaluate the performance of CodeBERT for detecting vulnerable functions in industrial and open-source software. We analyse its cross-domain generalisation when fine-tuned on open-source data and tested on industrial data, and vice versa, also exploring strategies for handling class imbalance. Based on these results, we develop \tool (\textit{A}utomating vulnerability detection \textit{I}ntegration for \textit{D}evelopers’ \textit{O}perations), a Continuous Integration–Continuous Deployment (CI/CD)-integrated recommender system that uses fine-tuned CodeBERT to detect and localise vulnerabilities during code review without disrupting workflows. Finally, we assess the tool’s perceived usefulness through a survey with the company’s IT professionals.
Our results show that models trained on industrial data detect vulnerabilities accurately within the same domain but lose performance on open-source code, while a deep learner fine-tuned on open data, with appropriate undersampling techniques, improves the detection of vulnerabilities.
\end{abstract}

\section{Introduction}
\label{sec:introduction}
The role of security experts in software development is of paramount importance. 
Their major task is to review developers' code pushed into a development pipeline and report back security weaknesses. Often, their work is performed manually as the last task before code delivery. They frequently rely on publicly available open-source information and, at times, use black-box static analysers that are not always integrated into the company’s development pipeline, creating custom rules to make the security scanner work for a specific application~\cite{Thomas2018Security}.

In this work, we aim to investigate whether a deep learner can be leveraged to enhance the automatic detection of vulnerable functions in industrial software and how it can be integrated into a company's DevOps environment. To this end, we first evaluated the performance of CodeBERT, a deep learner pre-trained on source code on the detection of vulnerable functions in both industrial and open-source software. Then, we examined whether this capability is retained when the model is fine-tuned on open-source data and then tested on industrial data.
We further developed a tool (\tool) to automate the detection of vulnerable functions in an industrial DevOps environment embedding a deep learner, it is designed to allow to exchange the deep learner for future adjustments. \tool~has been developed as a recommender system running over the Continuous Integration - Continuous Deployment (CI/CD) pipeline with a deep learner pre-trained on open source. It leverages multi-channel monitoring data to automatically detect and localise security weaknesses in code and present them to reviewers in their native environment. 
Finally, we gathered feedback from the company’s IT professionals regarding the tool. 
To guide the analysis for our case study, we have defined two research questions:

\newcommand{\RQone}{\textbf{RQ1.} \textit{Can CodeBERT  be employed  for vulnerability detection in industry source code?
}}
\noindent\RQone~This research question aims to explore how a deep learning model can be leveraged for the detection of vulnerabilities in industry. Our approach is therefore twofold: first, we compare the performance of a pre-trained model fine-tuned and tested with different balancing strategies on industrial data and open source data.  
Then, we develop \tool, embed it into the industrial DevOps of a company and ask the opinion of the company's IT professionals. 

\newcommand{\RQtwo}{\textbf{RQ2.} \textit{How well does CodeBERT fine-tuned on open source vulnerability data generalise to industrial technology-specific data?}}
\noindent\RQtwo~
This research question investigates the performance of fine-tuned models trained on open data and tested against industry data and vice versa. Furthermore, we explored various strategies for fine-tuning to mitigate data imbalance.

Overall, the contribution can be summarized with the following:
\begin{itemize}
    \item We performed a cross-domain performance evaluation on between datasets collected in industry and open data and made them available.
    \item We implemented \tool, which supports reviewers in detecting  vulnerabilities during the review process; furthermore, we have surveyed developer's about the a tool, \eg \tool, can support the review process.
\end{itemize}
\noindent The paper is structured as follows: \secref{sec:relatedWork} contains the related work, followed by the methodology in \secref{sec:methodologoy}.\secref{sec:caseStudy} presents the case study, going over to the \secref{sec:threats} and \ref{sec:conclusion} which discuss the the threats to validity and conclusion.
\section{Related Work}
\label{sec:relatedWork}
We have reviewed the existing literature according to two dimensions: (i) AI models for vulnerability detection, and (ii) cross-domain evaluation of datasets collected in open-source and industry contexts.
\subsection{AI Models for Vulnerability Detection}
\label{sec:AImodels}
Vaswani \etal~\cite{VaswaniEtAl2017} introduced the transformer architecture, enabling efficient learning across diverse tasks. Typically, transformer models are pre-trained on general tasks and subsequently fine-tuned for specific applications, such as vulnerability detection in source code. BERT~\cite{DevelinEtAl2019BERT}, trained on an English corpus, was later extended by RoBERTa~\cite{ChenEtAl2020RoBERTa}, which modified the training data, objectives, and parameters, while still focusing on the English language. Building on previous developments, CodeBERT~\cite{feng2020codebert} leverages multiple programming languages into its pre-training, using the CodeSearchNet dataset~\cite{HusainEtAl2019Codesearchnet}. This domain-specific training improves performance on code-related tasks compared to general-purpose models such as RoBERTa.
Fu \etal~\cite{fu2022linevul} proposed LineVul, a model designed to detect vulnerabilities using the Big-Vul dataset~\cite{FanEtAl2020BigVul}, which is derived from CVE-referenced commits. Their approach initially focused on function-level vulnerability detection and was later extended to perform detections at the line level. In their work, Fu \etal employed CodeBERT as the underlying model and processed source code at the function level, treating it as plain text without incorporating any structural or positional information, later their approach was refined for line-level detection. 
Hin \etal~\cite{HinEtAl2022LineVD} introduced LineVD, a deep learning framework, leveraging CodeBERT, that reframes statement-level vulnerability detection as a node classification task over program dependency graphs. By combining graph neural networks—capable of capturing control- and data-dependency structures—with to process raw source tokens, LineVD significantly outperforms prior techniques, achieving over a 105\% improvement in F1-score on Big-Vul, a real-world C/C++ vulnerability dataset. 
\subsection{Cross-Domain Performance Evaluation of Datasets}
\label{sec:cross}
Evaluating the performance of a novel approach is often limited to a single dataset, which may have been specifically created for that particular approach. However, assessing a tool's performance is resource- and time-consuming and using different benchmarks can unveil a lack of robustness in the approach.
A comparison of different Static Application Security Testing (SAST) tools, focusing on their performance in detecting vulnerabilities in Java projects indicating that the combination of multiple SAST tools increases the accuracy of detecting vulnerabilities \cite{LiEtAl2023ComparisonSASTJava}; however, it is not considered to solely relay on them. In contrast, when evaluating different deep learning approaches on a new created dataset, trained and tested, resulted into a performance drop of up to 91\% \cite{ChakrabortyEtAl2024ComparingDL}, indicating that a shift in the dataset makes highly specialized deep learning techniques impractical.

Despite the growing need to protect industry code from vulnerabilities \cite{DongEtAl2024IndustryIssue,PanEtAl2024OSSIndustry}, to the best of our knowledge, no cross-domain evaluation has been performed between open data, which are widely used in academia, and industry data. Therefore, the evaluation conducted in this work represents a critical first step in bridging the gap between academic research and industry practices. In contrast to the existing work, we do not focus solely on open data but investigate the performance of the a deep learner, in our case CodeBERT \cite{feng2020codebert}, trained on open data and applied on industry data. For the particular use case of the company, as an integration within the review process of source code.
\section{Methodology}
\label{sec:methodologoy}
This section illustrates our methodology, starting with the creation of the dataset, going over to the multiple fine-tuning strategies of CodeBERT and cross-validation, concluding with the implementation details. 
\subsection{Dataset creation and annotation}
\label{sec:dataset}
To the best of our knowledge, there are no publicly available datasets of PHP functions annotated for vulnerability using a consistent methodology for both open and industrial code.
As such, in this section, we present our technique to create and fuse three datasets: one from industry data, \ie \textit{Industry Dataset} (\ID) and the other two from open source data, the former from popular technology-agnostic open source projects, \ie \textit{Generic Open-source Dataset} (\GOD), and the latter from open source data originated from the same type of technology of the industrial data, \ie \textit{Technology-similar Open-source Dataset} (\TOD).

\begin{figure*}[t!]
    \centering
\includegraphics[width=\textwidth]{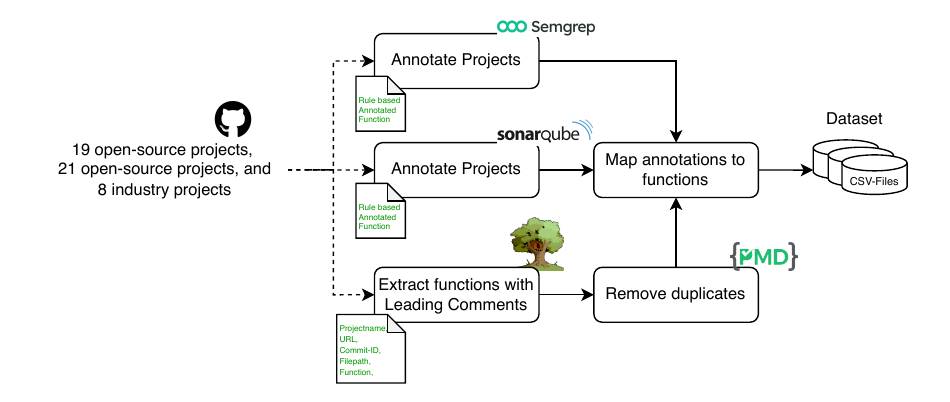}
    \caption{Creation of an annotated dataset by data fusion of two annotation types from static analysers for 40 open source and 8 industry projects.}
    \label{fig:dataset}
\end{figure*}

For each of the datasets, we applied  the same annotation process as illustrated in \figref{fig:dataset}.
To extract the PHP functions, we leveraged TreeSitter \cite{treesitter}, which was already employed successfully in recent mining studies \cite{MockEtAl2024Concerns,MockEtAl2024Dataset}. 
After extracting the functions from each project, we removed duplicates. We used PMD-CPD~\cite{pmd}, which detects duplicated code using a sliding-window approach and 99\% Jaccard index threshold~\cite{Sneath1957JaccardIndex}. Following Mock \etal~\cite{MockEtAl2024Dataset}, we set the window size to 30, to balance coverage and computational cost. 

Combining annotations from different tools helps detect a greater number of vulnerabilities in source code~\cite{NguyenDucEtAl2021CombiningSAST}, we performed function-level annotations at the project level and mapped the output to the extracted functions. We leveraged two state-of-the-art static analysers—SemGrep~\cite{semgrep} and SonarQube~\cite{sonarqube}—which are widely used and accepted in both industry and academia~\cite{LenarduzziEtAl2020SonarQube,ImprotaEtAl2025Semgrep}. We reviewed the resulting function annotations and decided not to leverage the labels Info or Minor from SemGrep, and Warning from SonarQube, as indicators of vulnerabilities. The annotation is counted uniquely, \ie a function might have multiple vulnerabilities; however it is counted once.
The schema of the datasets consists of two main parts. The first part contains  the information that helps to extract a function, \ie the url, commit-id, filename, and position of the function in the original file.  
The second part contains the annotation per severity level. The data is provided in three distinct CSV files, one for each of the datasets.

The datasets were created to meet the quality standards defined by the framework proposed by Croft \etal~\cite{CroftEtAl2023DataQuality}, which specifies five attributes: accuracy, uniqueness, consistency, completeness, and currentness. 
\textit{Accuracy} refers to the extent to which the data correctly represents the true value of the intended property. In our context, this means whether each function is accurately labelled as vulnerable or not. We employed Semgrep and SonarQube, two state-of-the-art SAST tools, for the automatic labelling of the functions.
\textit{Uniqueness} is the degree to which there is no duplication in records. In our context, this means whether each function has been unique in the dataset. We leveraged PMD-CPD \cite{pmd} with 30-token sliding window to detect duplicate code snippets across functions. To remove ducplicates, we also compute the Jaccard similarity of each function pair and removed functions of 99\% or higher similarity.
\textit{Consistency} is the degree to which data has attributes that are free from contradiction and are coherent with other data. In our context, this means whether each function have been annotated in a consistent manner with the different analysers over the different types of datasets. For this we leveraged the two SAST tools in the same version and setting across annotation of the three datasets.
\textit{Completeness} refers to the extent to which the data associated with an entity contains values for all expected attributes and related instances. In our context, this means that (1) the dataset  must include sufficient information for each function to be uniquely identified within the source code and (2) the output of the static analysers can be used to clearly identify a function as vulnerable or not. For this, we include the Url, commit-ID, file path, and start/end-position of each functions using TreeSitter \cite{treesitter}, a light-weighted code parser, such that the function can be traced back to the original location\footnote{Due to company restrictions, the Url and commit ID are omitted for the dataset \ID}. The output of the SAST tools is mapped automatically to the associated functions and stored by severity level; the scripts are available in the replication package \cite{MockEtAl2025replicationpackage}.
\textit{Currentness} is the degree to which data has attributes that are of the right age. 
In our context, this means that the data of the open source datasets pertain to the same period of time of industrial dataset. All datasets were created simultaneously leveraging the same SAST tools and configurations, ensuring temporal alignment of vulnerabilities.
\subsection{CodeBERT fine-tuning}
\label{sec:codeBERT}
CodeBERT is a pre-trained deep learner on six programming languages developed by Microsoft Research \cite{feng2020codebert}. It is designed specifically to work with both natural and programming language inputs.
CodeBERT has been trained on the dataset CodeSearchNet \cite{HusainEtAl2019Codesearchnet}. CodeSearchNet covers six programming languages (Python, Java, JavaScript, PHP, Ruby, and Go) and provides general code understanding but lacks vulnerability information. CodeBERT has better performance than other pre-trained models, \eg UniXcoder \cite{GouEtal2022UniXcoder}, also in different software engineering settings \cite{MockEtAl2025NLBSE}.

Fine-tuning refers to the process of further training a pre-trained model on a specific downstream task, allowing it to adapt to task-specific data and improve performance. During this process, the weights of the models are changed so that the model can perform the desired task \cite{DevelinEtAl2019BERT}. 
In our case, we performed fine-tuning on the pre-trained model CodeBERT~\cite{feng2020codebert}. 
We have fine-tuned CodeBERT on each of the three datasets for which we set the block size, batch size, and learning rate as 512, 16 and 2e-5 respectively, as suggested in the recent literature \cite{fu2022linevul,RussoEtAl2025VulSATD}. We fine-tuned each model for 10 epochs and performed a manual early stopping \cite{prechelt2002early} if the F1 increases more than 0.001 within five epochs. 
We randomly split each dataset in 80\%/10\%/10\%, for training, validation, and testing respectively. Furthermore, we trained and validated the model using four different data balancing techniques: (NB) no balancing, \ie the natural distribution of the dataset is taken; (USC) undersampling with the size of the smallest class across all datasets \cite{LeEtAl2024DataImbalanceSmallestClass}, \ie in our case, the vulnerable instances of \ID constitute the smallest class with 4{,}934 instances, and all other classes have been balanced accordingly; (URSC) undersampling with the size of the relatively smallest class per dataset \cite{fu2022linevul}, \ie in each dataset, the smaller class (the vulnerable class) determines the sample size for the non-vulnerable class; and (WLF) weighted loss function with inverse class frequency \cite{RussoEtAl2025VulSATD}, \ie the natural distribution of the dataset is maintained, but during training, the importance of the two classes is differentiated by adjusting the loss function’s weights. The loss function quantifies the discrepancy between the model's detections and the actual values during training, serving as a metric to guide optimization by penalizing inaccurate detections \cite{MockEtAl2025NLBSE}. No balancing was applied to the validation and testing set, \ie the natural distribution of the datasets were maintained.
\subsection{Cross-Domain Evaluation}\label{sec:crossdomain}
We used precision, recall, and the F1 score to evaluate the performance of CodeBERT in vulnerability detection in the testing portion of the three data sets. We repeat the analysis for each of the balancing techniques. Performance measures are computed as follows: 
\begin{equation}
    \label{eq:p}
    \text{P} = \frac{TP}{TP + FP}, \,\text{R} = \frac{TP}{TP + FN}, \, \text{F1} = 2 \cdot \frac{P \cdot R}{P + R}.
\end{equation}
The goal of cross-domain analysis in our context is to evaluate the performance of CodeBERT on industry data after it has been fine-tuned on open source data. This analysis helps answer questions such as: \RQone and \RQtwo~To investigate these problems, we fine-tune CodeBERT on the two datasets \GOD and \TOD and analyse its performance on the testing portion of the \ID dataset.
We then compared CodeBERT's performance on \ID with its performance on the same dataset on which it has been fine-tuned, also using different balancing strategies.
\section{Case Study}
\label{sec:caseStudy}
The case study was performed at W{\"u}rth Phoenix, an Italian software company. We selected an ERP project written in PHP as it is representative of the company's business. The project has over 2.2 million lines of code, spanning over 74 thousand commits written in a period of up to 20 years. The development team consists of 15 members with the roles and IT experience reported in \tabref{tab:demographics}.

\begin{table}[t!]
    \centering
    \caption{Development team' role and IT experience}
    \label{tab:demographics}
    \setlength{\tabcolsep}{7pt}
    \begin{tabular}{lcc}
    \hline
    \textbf{Role} & \textbf{\#} & \textbf{Years working in industry}   \\ \hline
    Software Developer & 6 & 1-3 years  \\ 
    Software Engineer & 2 &  12, 15 years  \\
    Software Engineer in R\&D & 1 & 2 years  \\
    Security Software Engineer & 1 & 4 years  \\
    Software Architect & 2 & 5, 11 years  \\
    DevOps & 2 & 5 years  \\
    Team Leader & 1 & 15 years  \\ \hline
    \end{tabular}
\end{table}
The company's internal CI/CD pipeline runs on a self-hosted and maintained instance of OpenShift \cite{openshift}.
Before the release of every new feature of the project, a five-step workflow is sequentially followed in the pipeline: low-level code design, development, code review, security review, and staging.
\textit{Low-level code design.}
When a new feature is requested by a customer or internally, a low-level code design is performed~\cite{Gamma95Reusable,McConnell2004Code}.
Multiple developers internally discuss the new feature, including how it can be implemented, where best to place it in the architecture of the current source code, and, if the current code needs to be modified, how to do this with minimal impact while still maintaining high reusability for the future.
If needed, the feature is broken down into multiple tasks so that it is easier to handle them within a sprint~\cite{SchwaberEtAl2011Sprint}. The security expert participates in Sprint meetings to anticipate potential security risks and ensure security by design.
\textit{Development.}
Based on the JIRA issues, developers implement the requested feature. A single task can be a feature or part of a larger feature. At this stage, no reviewers (code or security) are involved.
\textit{Code Review.}
In this stage, newly added, changed, or removed code is inspected regarding architectural consequences, maintainability, and general quality of the code. Developers, architects, and software engineers are typically involved.
\textit{Security Review.}
In the security review, newly added, changed, or removed source code is inspected for potential vulnerabilities, following the same schema as the code review by the security expert.
\textit{Deployment.}
Issues' resolutions follow a pre-defined deployment schedule if they are general features. However, if they are vulnerability patches, they are deployed directly without respecting the general schedule.
\subsection{Data collection}
\label{sec:dataCollection}
\begin{table}[tb]
    \centering
    \caption{
    Summary of annotations categorised by severity levels and tools. 
    }
    \label{tab:vulnerabilities}
    \setlength{\tabcolsep}{8pt}
    \begin{tabular}{p{0.9cm}P{0.9cm}P{2.4cm}|ccc|P{1cm}}
    \hline
    &&& \multicolumn{3}{c|}{Sonarqube}  & Semgrep\\ 
        Dataset & \# fns & \# vulnerable fns&Major  & Critical & Blocker     & Error   \\ \hline
        \GOD   & 206,647   &48,797      & 38,249 & 16,084   & 2,061       & 647     \\
        \TOD & 238,161   &21,579      & 16,210 & 10,195   & 2,361       & 122      \\
        \ID   & 64,333  &6,149        & 3,337  & 4,094    & 322         & 0       \\ \hline
        Total &509,141	& 76,525&57,796	&30,373	&4,744	&769 \\ \hline
    \end{tabular}
\end{table}
The dataset \GOD contains 209,532 functions of 21 open-source PHP projects selected based on their popularity. The dataset \TOD contains 240,876 functions of the top 19 GitHub Enterprise Resource Planning tools (\emph{ERP}) written in PHP (forks have been excluded). The dataset \ID consists of 65,385 functions of eight ERP projects of the partner company. For each project, we collected all the functions included in the most recent version of project's repository.
We then applied our approach and successfully removed 6,136 duplicated functions (1,052 in \ID, 2,885 in \GOD, and 2,229 in \TOD), which were consistently duplicated with the two SAST tools.
76,181 were marked as vulnerable by SonarQube, and an additional 769 by SemGrep, resulting in a combined total of 76,525 vulnerable functions. It should be noted that the sum does not add up due to partial overlaps of the different severity levels.
\tabref{tab:vulnerabilities} illustrates the resulting datasets, with further insights into the severity levels from each SAST tool.
\subsection{Cross-Domain Performance Evaluation}
\label{sec:CDPerformanceEvaluation}
\begin{figure}[t!]
    \centering
    \begin{subfigure}[b]{\textwidth}
        \centering
 \includegraphics[width=\textwidth]{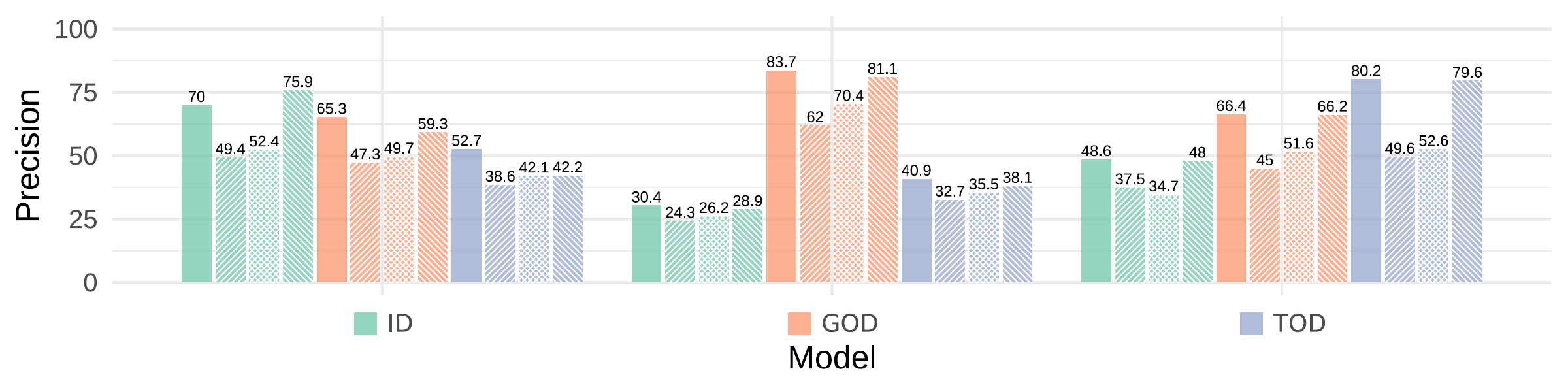}
        \label{fig:performanceEval_precision}
    \end{subfigure}
    \\
    \begin{subfigure}[b]{\textwidth}
        \centering
        \includegraphics[width=\textwidth]{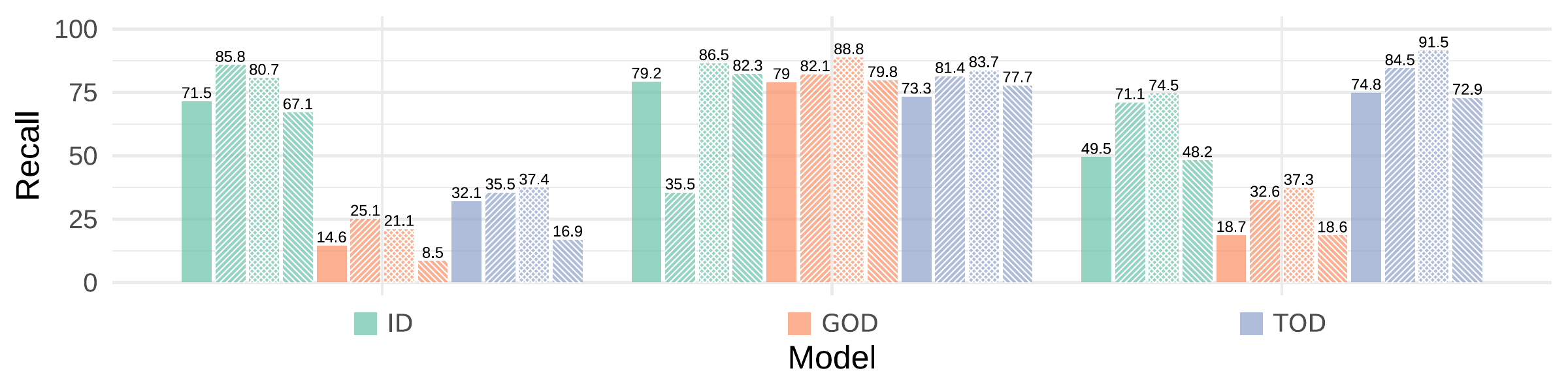}
        \label{fig:performanceEval_recall}
    \end{subfigure}
    \\
    \begin{subfigure}[b]{\textwidth}
        \centering
        \includegraphics[width=\textwidth]{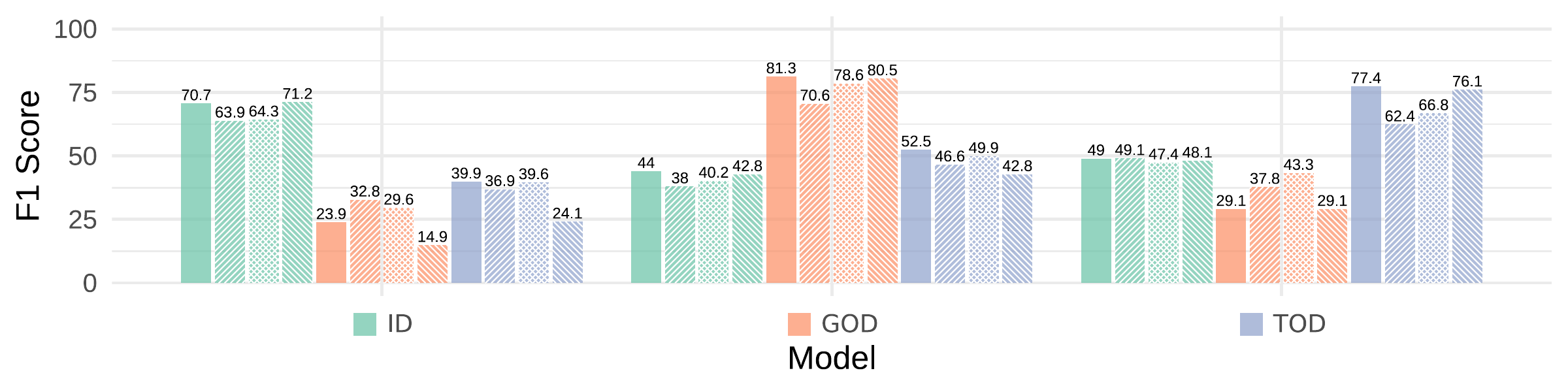}
        \label{fig:performanceEval_F1}
    \end{subfigure}
    \\
    \begin{subfigure}[b]{\textwidth}
        \centering
        \includegraphics[width=\textwidth]{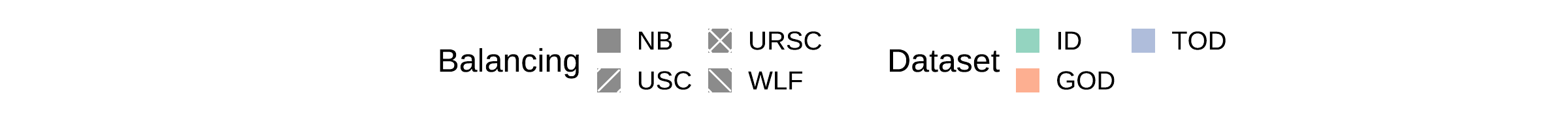}
    \end{subfigure}
    \caption{Testing performance of CodeBERT fine-tuned on different datasets and with different balancing strategies.}
    \label{fig:performance}
\end{figure}
Plots in \figref{fig:performance} illustrate Precision, Recall, and the F1 score of CodeBERT in each of the configurations of this study: the colour of the bars represents the type of dataset chosen for testing. For example, green means that the performance of CodeBERT has been tested on \ID. The pattern of the bars represents the balancing strategy. For instance, no pattern means that fine-tuning has been performed without balancing. On the x-axis, the labels indicate the datasets which have been used to fine-tune CodeBERT, \ie trained and validated on the vulnerability task. Thus, the first green bar in \figref{fig:performance} refers to the F1 score (70.7) of CodeBERT that has been fine-tuned without balancing and tested on \ID.

In vulnerability detection, the primary objective is to minimize false negatives (FN), since even a single undetected vulnerable function can lead to significant damage if exploited~\cite{LiEtAl2018vuldeepecker}. Therefore, in the following analysis, we  prioritize recall over precision, as capturing all potential vulnerabilities is considered more critical than avoiding false alarms.
To guide our analysis, we applied a threefold approach: (1) analyse the most effective balancing strategy, (2) analyse how fine-tuning on different datasets affects the performance of CodeBERT on unseen data, and (3) restrict our analysis to the performance of CodeBERT on industrial unseen data. 

To analyse the most effective balancing strategy, we analyse the plots in terms of the four balancing strategies described in \secref{sec:codeBERT}.
\begin{itemize}
\item  When both fine-tuning and testing are performed with the same dataset, we observe higher F1 score and precision with NB or WLF, whereas any type of \textit{undersampling} is generally preferred to reduce FNs (increased recall).  
\item When fine-tuning is performed with \ID and \TOD and testing with \GOD, we observe higher F1 score and recall with any type of \textit{undersampling} techniques, whereas  NB or WLF are better choices to increase precision. 
\item When fine-tuning is performed with \ID and \GOD and testing with \TOD, we observe higher F1 score and precision without balancing (NB), whereas any type of \textit{undersampling} techniques  is preferred to increase recall.
\item When fine-tuning is performed on \TOD and \GOD and testing is computed on \ID, we observe higher F1 score and precision with NB and WLF, whereas  any type of \textit{undersampling} techniques  is preferred to increase recall.
\end{itemize}

In summary, when maximizing recall is the primary objective, we recommend applying undersampling techniques. Conversely, for optimizing overall performance as measured by the F1-score, either no balancing or the use of a weighted loss function is more appropriate.

We then study how fine-tuning on different datasets affects the performance of CodeBERT on unseen data. 
\begin{itemize}
    \item  When CodeBERT has been fine-tuned with \TOD, we observe the best performance for all measures  is testing with the same dataset. In addition, recall is better when testing with \ID than with the dataset \GOD.
    \item When CodeBERT has been fine-tuned with \GOD, we observe the best performance for Precision and F1 score when testing with \GOD, whereas testing with \ID produces similar - sometimes even better - recall. 
    \item When CodeBERT has been fine-tuned with \ID, we observe the best performance for all measures when testing with \ID itself. Testing with open source data produce a poor recall.
\end{itemize}
Generally speaking, testing on the same dataset on which CodeBERT has been fine-tuned yields better performance. However, in terms of FNs,  we can observe that (1)  Fine-tuning on open-source data and testing on industrial data — both from the same technology - results in fewer FNs compared to testing on general open-source data.  Technology-related open source data does not generalize well to general open source data. 
(2) Fine-tuning on industrial data and testing on open-source data leads to a significant increase of FNs. 
(3) Fine-tuning on technology-agnostic open source data is worth to detect vulnerabilities of industrial code (similar or better recall).
Motivated by the last result, we pose the following question.
Thus, when we further restrict  our analysis to the performance of CodeBERT on industrial unseen data we notice that:
\begin{itemize}
\item  precision and F1 score decrease if CodeBERT has been fine-tuned with any open-source data; 
\item fine-tuning with \GOD combined with undersampling within the same dataset (URSC) yields higher recall than fine-tuning with \ID or \TOD using any balancing techniques. In addition,  CodeBERT achieves higher recall if fine-tuned on the \GOD dataset rather than on the \ID dataset, whether using NB or WLF techniques.
\end{itemize}

In summary, using technology-agnostic open-source data to detect unseen industrial vulnerabilities can result in better control of FNs than using industrial data from the same system. However, FPs are an issue. 
\subsection{\tool}
\label{sec:tool}
\begin{figure*}[t!]
    \centering
    \includegraphics[width=0.9\textwidth]{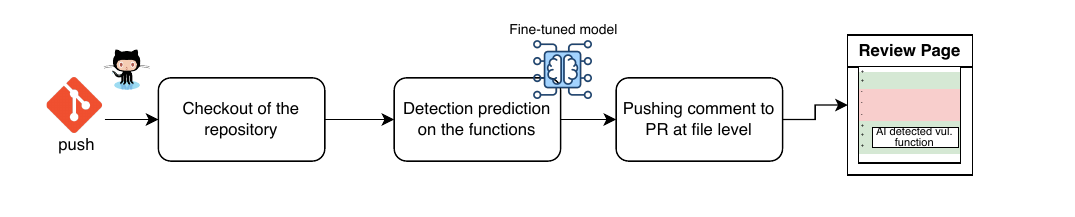}
    \caption{Flow of the pipeline integrated into a GitHub Action.}
    \label{fig:flow}
\end{figure*}
In the previous section, we showed that CodeBERT successfully detects vulnerabilities in industrial software. We also observed that CodeBERT can be fine-tuned on technology-agnostic open source data for this task. 
Thus, we developed \tool that leverages an instance of CodeBERT fine-tuned on \GOD on the DevOps pipeline of the company illustrated in \secref{sec:caseStudy}. 
\tool~is integrated within a GitHub Action; furthermore, it leverages git \cite{git} and TreeSitter \cite{treesitter} for the automatic extraction of functions changed or newly added within a pull request. \figref{fig:flow} illustrates the process of pushing to the repository, triggering the pipeline, detecting the vulnerabilities, and commenting on the pull request (PR).
The first step is divided into five sub-steps critical to the vulnerability detection process. First, git diff identifies the affected file paths and the corresponding modified lines.
The repository is checked out with both the \texttt{destination} and \texttt{source} branches.
The files impacted by the pull request are extracted using git diff, along with the modified line ranges.
Functions from these files are parsed using TreeSitter~\cite{treesitter} and filtered based on their overlap with the modified lines to ensure relevance.
The fine-tuned instance of CodeBERT is applied to the relevant functions. 
Functions detected as vulnerable are automatically tagged to support reviewer decision-making. An example setup and the source code are available in the replication package~\cite{MockEtAl2025replicationpackage}.
Finally, we presented \tool to the IT professionals of the company and collected their opinions with a brief survey. 

\tool hosts a deep learner that is fine-tuned offline, ensuring complete non-intrusiveness in the development process. This model can be periodically updated offline with new data to improve accuracy, if necessary. Furthermore, \tool is designed to allow the fine-tuned model to be replaced, enhancing its usability and applicability beyond the scope of our work.

\noindent\textbf{Developers' Perspective}
We administered a brief questionnaire about \tool \footnote{Available in the replication package \cite{MockEtAl2025replicationpackage}.} to 13 of the 15 IT professionals at the company (two developers were absent).  
The questionnaire collects information about their review process, their opinion on a line-level vulnerability detection tool like \tool, and their demographic. The questionnaire alternates closed and open questions. 

We review their answers by their role as illustrate in \tabref{tab:demographics}.
The security expert has been conducting code reviews focused on security for four years, multiple times per week, identifying vulnerabilities several times each month. However, he is only moderately enthusiastic about the tool. This is a typical reaction among specialized professionals, who often believe that new tools might eventually put at danger their role in the organization~\cite{FitzgeraldEtal2011AdoptingOpenSource}.
The two DevOps experts have different levels of experience with vulnerability detection during the review process. Both review code weekly to identify bugs; however, the expert who encounters vulnerabilities more frequently—approximately one per month—expressed greater appreciation for the tool, stating: “This tool is really useful to speed up and improve the development process! It could help to spot issues before they go to production.” Conversely, the expert who rarely encounters vulnerabilities is the most negative among all participants and suggests that the tool should be enhanced with a more extensive set of detection rules.
The three software engineers, although they do not normally find many vulnerabilities during their review tasks, greatly appreciate the tool. Two of them have extensive experience in IT and code review. The third is a junior engineer whose work focuses more on process quality than on code, as she rarely performs code reviews.
The four developers are junior professionals who perform code reviews frequently—multiple times a week—as it is one of their primary tasks. All are moderately satisfied with the tool, although they acknowledge lacking sufficient experience in vulnerability detection.
The two software architects, who review code as frequently as the developers, are more enthusiastic about the tool. They both recommend enhancing it to provide more detailed information on the type of vulnerability and to explain the reasoning behind AI-DO’s output (\ie the vulnerable lines of code).
The team leader is an experienced professional who performs code reviews sporadically, approximately once per month, also with the specific aim of detecting vulnerabilities. He moderately appreciate  the tool, considering it insufficiently precise, as vulnerabilities often span multiple lines and classes.

In summary, professionals’ enthusiasm for security tools largely depends on their experience with vulnerability detection and how well the tool supports their specific needs. While experts who frequently find vulnerabilities see clear benefits, those with less exposure or more specialized roles tend to be more cautious or critical, often concerned about the tool’s limitations or its potential impact on their expertise.

\noindent\textbf{Implementation Details and Replication Package}
The implementation details can be found in the replication package \cite{MockEtAl2025replicationpackage}.

\subsection{Answers to the Research Questions}
\label{sec:answerRQ}

\RQone~Yes, CodeBERT fine-tuned on  data of the company performs comparatively well on industrial unseen data with at most a 10\% decrease of the F1 score with respect to CodeBERT fine-tuned and tested on open source unseen data. Using a weighted loss function in fine tuning this gaps further reduces. 
However, when the focus is on reducing FNs, undersampling techniques can be more efficacious.
After introducing \tool, which implements our approach in the company’s DevOps PHP environment, we observed that professionals’ enthusiasm for the tool  depends on their experience with vulnerability detection and the extent to which it addresses their specific needs. Experts who frequently detect vulnerabilities recognize clear benefits, whereas those with less exposure or highly specialized roles are more cautious as they may see its potential to diminish the value of their expertise.

\begin{mdframed}[backgroundcolor=gray!10]
CodeBERT is suitable for vulnerability detection in the industrial domain, achieving performance comparable to its detection capabilities in the open-source domain.  Undersampling techniques can help decrease the number of FNs in both domain. Automating vulnerability detection in an industrial DevOps environment characterized by clearly defined roles   eventually needs to address the resistance of highly specialized professionals who may perceive their roles as being at risk.
\end{mdframed}

\RQtwo~We conducted a cross-domain evaluation by fine-tuning  CodeBERT  on technology-specific and technology-agnostic open-source data (resp. \TOD and \GOD) and testing it on unseen industrial data (\ID). 
We found that fine-tuning with URSC undersampling on a technology-agnostic open-source dataset can provide better control of false negatives in detecting unseen industrial vulnerabilities than fine-tuning industrial data from the same system with any balancing techniques.  This may be due to the greater heterogeneity of vulnerability types and coding styles present in the open-source dataset that can improve the model’s ability to identify real vulnerabilities in industrial systems, reducing false negatives that could otherwise leave security flaws unaddressed. However, this detection capability comes at the cost of increased false positives, which can burden development teams with unnecessary investigations and potentially reduce trust in \tool.

\begin{mdframed}[backgroundcolor=gray!10]
Fine-tuning CodeBERT with  technology-agnostic open-source vulnerability data using appropriate undersampling improves the detection of industrial vulnerabilities. However, this may increase professional' workload and affect trust in a detection tool. 
\end{mdframed}

\section{Threats to Validity}
\label{sec:threats}
\noindent\textit{Construct Validity: } Threats to construct validity refer to the extent to which the experimental setting actually reflects the construct under study.
We study the portability of a fine-tuned pre-trained deep learner for vulnerability detection between open and industry data. Three datasets were collected using the same; differences in sample size and vulnerability count were addressed by balancing and weighting strategies. The deep learner was created in the same way for all the datasets; enabling fair same-domain and cross-domain performance comparison.

\noindent\textit{Internal Validity:} Threats in this category are related to internal factors that could have influenced the results. 
Threats including duplicate code in the datasets, which we mitigated by PMD-CPD, and possible shared code between open and industry datasets from sources like Stack Overflow, which could not be identified. 

\noindent\textit{External Validity:} Threats in this category concern the generalizability of the results. 
Results may be specific to our datasets. A domain shift was found to reduce the generalizability, as intended in our study design. Results from the survey might be biased in favour of the work, as the first author had prior contact with the participants; a replication of the survey would strengthen the findings. PHP was selected as target programming language due to the need of the company for it, which might limit broader generalizability. 

\noindent\textit{Conclusion Validity:} This aspect regards the relationship between treatment and outcome, \ie if we can ascertain, with a given significance, that the outcome was a consequence of the treatment. 
We applied standard metrics to evaluate model performance to assess the performance of the deep learner.

\section{Conclusion and Further Work}
\label{sec:conclusion}
We evaluated a fine-tuned deep learning model for vulnerability detection in an industrial setting. The insights we collected led to the development of \tool, which we later evaluated by surveying developers from 13 companies to assess its value. To examine cross-domain applicability, we created three datasets annotated under the same policy: one company projects, and two from open-source projects—one domain-related and one general. Fine-tuning CodeBERT on the three datasets individually, and further assessing different balancing strategies, enabled us to study learning transfer.
Results show the approach is feasible and valuable for companies aiming to “shift left” vulnerability detection, \ie detecting vulnerable code as early in the development process as possible. Developers were open-minded regarding the tool, \tool, but concerns arose regarding trust and localisation of the solution. Models trained on industrial data performed well in-domain but poorly on general open-source code, likely due to differences in vulnerability types and coding standards. Training on technology-agnostic open data with undersampling increases the detection of vulnerabilities in the industrial dataset.
Future work includes extending to multiple programming languages, transformer models, and annotation types, leveraging our new multi-annotation dataset~\cite{MockEtAl2024Dataset}.
\begin{credits}
\subsubsection{\ackname} 
Funded by the European Union- Next Generation EU, Mission 4 Component 1 CUP I52B23000570003. 
The work has been funded by the project CyberSecurity Laboratory no. EFRE1039 under the 2023 EFRE/FESR program. We acknowledge ISCRA for awarding this project access to the LEONARDO supercomputer, owned by the EuroHPC Joint Undertaking, hosted by CINECA (Italy). We thank W{\"u}rth Phoenix for hosting the first author and providing the data for this~work. 
\end{credits}

\bibliographystyle{splncs04}
\bibliography{ref.bib}

\end{document}